\documentclass[conference]{IEEEtran}

\usepackage{amssymb} 	
\usepackage{color}
\usepackage{amsmath, bbm}
\usepackage{mathtools}
\usepackage{fullpage}
\usepackage[table,xcdraw]{xcolor}
\usepackage{multirow}
\usepackage[super]{nth}
\usepackage{graphicx,adjustbox}
\usepackage{caption}
\usepackage[labelformat=simple]{subcaption}
\usepackage{comment}
\usepackage{setspace}
\usepackage{textcomp}
\usepackage{xspace}
\usepackage{siunitx}
\usepackage{epsfig}
\usepackage{epstopdf}
\usepackage{soul}
\usepackage{url}
\usepackage{tablefootnote}
\usepackage[linesnumbered,ruled]{algorithm2e}
\usepackage{booktabs}
\usepackage[normalem]{ulem}
\usepackage{footnote}
\usepackage[misc,geometry]{ifsym} 
\makesavenoteenv{tabular}
\usepackage[utf8]{inputenc}
\usepackage{siunitx}
\usepackage[acronym,nomain,nonumberlist]{glossaries}
\usepackage{floatrow}
\usepackage{siunitx}
\newacronym{ap}{AP}{Access Point}
\newacronym{aiml}{AI/ML}{Artificial Intelligence and Machine Learning}
\newacronym{svm}{SVM}{Support Vector Machine}
\newacronym{arima}{ARIMA}{Autoregressive Integrated Moving Average}
\newacronym{ann}{ANN}{Artificial Neural Network}
\newacronym{rnn}{RNN}{Recurrent Neural Network}
\newacronym{lte}{LTE}{Long-Term Evolution}
\newacronym{mae}{MAE}{Mean Absolute Error}
\newacronym{mape}{MAPE}{Mean Absolute Percentage Error}
\newacronym{rmse}{RMSE}{Root Mean Squared Error}
\newacronym{mlp}{MLP}{Multi-Layer Perceptron}
\newacronym{sta}{STA}{STAtion}
\newacronym{lstm}{LSTM}{Long Short-Term Memory}
\newacronym{cnn}{CNN}{Convolutional Neural Network}
\newacronym{svr}{SVR}{Support Vector Regression}
\newacronym{gnn}{GNN}{Graph Neural Network}
\newacronym{mac}{MAC}{Medium Access Control}
\newacronym{cdf}{CDF}{Cumulative Distribution Function}
\DeclareSIUnit{\device}{device}
\newfloatcommand{capbtabbox}{table}[][\FBwidth]

\let\OldTexttrademark\texttrademark
\renewcommand{\texttrademark}{\OldTexttrademark\xspace}%

\usepackage{tikz}
\usepackage{tkz-tab}
\usetikzlibrary{automata,arrows,positioning,calc, fit, arrows.meta, }
\usetikzlibrary{shapes,snakes}
\usetikzlibrary{arrows}
\usepackage{multirow}
\usepackage{booktabs}
\usepackage{hyperref}
\usepackage{scalerel}
\pgfdeclarelayer{background}
\pgfdeclarelayer{foreground}
\pgfsetlayers{background,main,foreground}

\usepackage{scalefnt}
\usepackage{subfiles}
\usepackage{neuralnetwork}

\usepackage{booktabs}
\usepackage{multirow}

\def\BibTeX{{\rm B\kern-.05em{\sc i\kern-.025em b}\kern-.08em
    T\kern-.1667em\lower.7ex\hbox{E}\kern-.125emX}}

\begin{document}


\title{AI/ML-based Load Prediction in IEEE 802.11 Enterprise Networks}

\author{
\IEEEauthorblockN{Francesc Wilhelmi$^{\flat}$, Dariush Salami$^{\star}$, Gianluca Fontanesi$^{\flat}$, Lorenzo Galati-Giordano$^{\flat}$, and Mika Kasslin$^{\star}$ \vspace{0.1cm}
}
\IEEEauthorblockA{$^{\flat}$\emph{Radio Systems Research, Nokia Bell Labs, Stuttgart, Germany}}
\IEEEauthorblockA{$^{\star}$\emph{Radio Systems Research, Nokia Bell Labs, Espoo, Finland}}
}

\maketitle

\begin{abstract}
Enterprise Wi-Fi networks can greatly benefit from \gls{aiml} thanks to their well-developed management and operation capabilities. At the same time, \gls{aiml}-based traffic/load prediction is one of the most appealing data-driven solutions to improve the Wi-Fi experience, either through the enablement of autonomous operation or by boosting troubleshooting with forecasted network utilization. In this paper, we study the suitability and feasibility of adopting \gls{aiml}-based load prediction in practical enterprise Wi-Fi networks. While leveraging \gls{aiml} solutions can potentially contribute to optimizing Wi-Fi networks in terms of energy efficiency, performance, and reliability, their effective adoption is constrained to aspects like data availability and quality, computational capabilities, and energy consumption. Our results show that hardware-constrained AI/ML models can potentially predict network load with less than 20\% average error and 3\% 85th-percentile error, which constitutes a suitable input for proactively driving Wi-Fi network optimization.
\end{abstract}



\section{Introduction}
\label{section:introduction}

Generation after generation Wi-Fi is including new capabilities to embrace novel use cases like immersive communications, digital twins for manufacturing, or cooperative robotics~\cite{galati2023will}. These use cases come along with ever-increasing and expanded network requirements such as ultra-low-latency, superior reliability, and mobility~\cite{khorsandi2022hexa}. Such stringent requirements demand extremely fast-responding network management and operation, able to flexibly adapt to rapidly varying situations. And to that end, Artificial Intelligence and Machine Learning (AI/ML)-based traffic prediction stands as an appealing tool for driving autonomous network optimization~\cite{troia2018deep} by unlocking key applications like proactive mobility management, resource allocation, or energy efficiency. 

Enterprise and industrial Wi-Fi networks can benefit the most from \gls{aiml}-based traffic prediction thanks to a centrally managed and operated infrastructure. While the IEEE 802.11 \gls{aiml} Task Interest Group (TIG)~\cite{aiml_tig} has just started to discuss relevant use cases and potential standard implications for \gls{aiml} in next-generation Wi-Fi, enterprise and industrial Wi-Fi solutions are nowadays mainly relying on \gls{aiml}-based proprietary solutions. In operator-managed Wi-Fi networks, \gls{aiml} procedures such as data collection and ML-driven network configuration are typically enforced by existing standards such as the Wi-Fi CERTIFIED Data Elements~\cite{wfa_dataelem}.

Several studies~\cite{tedjopurnomo2020survey,abbasi2021deep} have shown remarkable performance in several enterprise Wi-Fi deployments by leveraging \gls{aiml}-based traffic prediction solutions. However, the practical adoption of \gls{aiml} solutions still raises concerns and leaves many questions unresolved. The complex nature of events occurring in networks at different time scales (e.g., from \unit{\milli\second}-order variations to yearly-seasonal patterns) poses a trade-off between the achievable performance by \gls{aiml} methods and their complexity, which entails costs in terms of computation and memory needs, energy, and time.

In this paper, we study the gap between the performance of relevant state-of-the-art \gls{aiml} models and the cost and feasibility of training and maintaining them in practice. More specifically, we focus on traffic prediction for enterprise Wi-Fi utilizing an open-source dataset presented in \cite{chen2021flag}, which includes measurements from thousands \glspl{ap} in a wide campus network. Differently from the state-of-the-art, the challenge we address in this work is the resource trade-off in terms of data collection, training and re-training needs, computational complexity, and energy when designing ML models to be deployed in real environments.

The rest of the paper is structured as follows: Section~\ref{section:load_prediction} overviews the related work on load prediction for wireless networks. Then, Section~\ref{section:methodology} defines the time series forecasting problem through the angle of \gls{aiml} and describes both data and models used for evaluation. Section~\ref{section:evaluation} presents the main results of our work and Section~\ref{section:conclusions} concludes the paper with final remarks.

\section{Related work}
\label{section:load_prediction}

Network traffic prediction has been largely considered as a time series forecasting problem~\cite{kirchgassner2012introduction} whereby the future values of a signal are predicted by leveraging temporal correlations. The \gls{arima} model, one of the most used time series forecasting techniques, was adopted in~\cite{hernandez2009arima} for Wi-Fi traffic prediction. However, given their simplicity, classical models like \gls{arima} fail to generalize and do not provide accurate predictions when the time series is complex, which is typically the case for network traffic embedding activities from a large number of users~\cite{jin2012characterizing}.

Conversely, \gls{aiml} holds the promise to cope with high-complexity data by automatically recognizing and learning patterns (e.g., complex spatial and temporal correlations). \gls{aiml} has been broadly applied to Wi-Fi-related problems~\cite{wilhelmi2020flexible, szott2022wi}. In traffic prediction, early works studied parametric methods such as \gls{svm}~\cite{feng2006svm} or polynomial regression~\cite{thapaliya2018predicting} to predict the future load or network data (e.g., average clients, throughput, and frame errors) in Wi-Fi deployments. Supervised learning techniques, compared to \gls{arima}, showed better performance when applied to real traffic data thanks to their noise-tolerance, stability, and adaptability characteristics. 

Other models such as \gls{ann}, \gls{svr}, decision trees, and random forests were studied in~\cite{khan2020real} for real-time throughput prediction in Wi-Fi. Furthermore, a more elaborated \gls{rnn}-based solution was proposed in~\cite{chen2021flag} for load prediction in a large campus Wi-Fi network. The \gls{rnn}-based solution in~\cite{chen2021flag}, thanks to its ability to learn spatial and temporal correlations, was shown to outperform traditional models like \gls{arima} and \gls{svm}.

Traffic prediction has also been extensively studied in cellular networks~\cite{jiang2022cellular}. In~\cite{trinh2018mobile}, \gls{lstm} was used to predict user load in \gls{lte} networks. \glspl{lstm} include specialized mechanisms to understand temporal sequences, which is useful to overcome the vanishing gradient problem of \glspl{rnn}. In contrast, \glspl{lstm} lack scalability and fail at capturing spatial interactions, as they are not designed for that purpose. An appealing alternative are \glspl{cnn}, which were popularized thanks to their incredible performance at recognizing images~\cite{albawi2017understanding}. The work in~\cite{zhang2020citywide} proposed a \gls{cnn}-based solution to leverage spatial correlations in mobile traffic, including SMS, voice, and Internet services. Another solution designed to capture spatial correlations from cellular sites was proposed in~\cite{feng2018deeptp}, which combined \gls{lstm}, \textit{seq2seq}, and self-attention to predict traffic load. Finally, the latest state-of-the-art models, including novel transformer architectures, were evaluated in~\cite{perifanis2023towards} for federated cellular traffic forecasting.

\section{Methodology}
\label{section:methodology}

\subsection{Time series forecasting}

Let $\mathbf{X}_{T-L:T}= \{\mathbf{x}_{T-L}, ..., \mathbf{x}_T\}$ be a series of features collected across $L$ consecutive \textit{lookback} points and up to time step $T$, where each measurement is a feature vector $\mathbf{x}_t\in \mathbb{R}^{M}$ containing $M$ features. The goal of the time series forecasting problem is to predict the values of a subset of selected features $N\subseteq M$ in the future next $S$ time \textit{steps}, i.e., $\hat{\mathbf{X}}_{T+1:S}$. For that, we resort to supervised learning, where available data is used to fit the weights $W$ of a forecasting function $f_W()$. In particular, the predicted values for the next $S$ steps are obtained as
\begin{equation}
    f_W(\mathbf{X}_{T-L+1:T}) = \hat{\mathbf{X}}_{T+1:S}.
\end{equation}

In this work, we use different models as forecasting function $f_W()$. The optimization of the models is driven by the minimization of a loss function $l()$, which compares each prediction with its ground truth:

\begin{equation}
    l(\hat{X}_{T+1:S}, X_{T+1:S}) = \{l_{T+1}, ..., l_S\}^T.
\end{equation}

\subsection{Evaluation metrics}

The performance of a model can be calculated as
\begin{equation}
A(f_W) = \sum_{p\in P} \epsilon_f(\mathcal{Y}_p,\hat{\mathcal{Y}}_p),
\end{equation}

where $\epsilon_f$ is an error function, and  $\hat{\mathcal{Y}}$ and $\mathcal{Y}$ are the considered predictions and actual values, respectively. In this work, we use the \gls{mape} as the main model accuracy metric, which is computed as
\begin{equation}
    \text{MAPE} =100\times \frac{1}{P} \sum_{p=1}^P \frac{|\mathcal{Y}_p - \hat{\mathcal{Y}}_p|}{|\mathcal{Y}_p|}.
\end{equation}

Although not upper-bounded, \gls{mape} provides a good intuition about the suitability of different models for assisting load-prediction-driven applications (e.g., switching off \glspl{ap} with a forecasted load below 10\%). Besides the prediction error, we also measure the energy consumption during training and inference. For that, we resort to \textit{Eco2ai}~\cite{budennyy2022eco2ai}, an open-source library to measure the energy consumed by ML models.

\subsection{Dataset}

We use the open-source dataset published in~\cite{chen2021flag}, which contains \num{25074733} association records (from a total of \num{55809} users) extracted from \num{7404} Wi-Fi \glspl{ap} distributed in a campus area of \qty{3}{\kilo\meter\squared}, during \num{49} days in \num{2019}. The data contains features such as the user \gls{mac} address, the ID of the \gls{ap} to which the user is associated, and the number of Bytes transferred per session (see~\cite{chen2021flag} for more details).

Taking the association records from the original dataset, we derived the temporal load (both uplink and downlink) of each \gls{ap} by aggregating the individual contributions from each client. In particular, we equally spread the total load $\rho^j$ of each connection $j$ by the number of time steps of size $w$ in the association duration $T^j$, where $w$ is a configurable parameter that defines the granularity of the measurements (e.g., \qty{1}{\second}, \qty{1}{\minute}, \qty{1}{\hour}).\footnote{For instance, an association record indicating that a station transfers \qty{100}{\mega\byte} during \qty{10}{\minute} is sampled into time steps of \qty{1}{\minute}, thus returning a data transfer of $100\text{MB}/(\qty{10}{\minute}/\qty{1}{\minute})=10$ \unit[per-mode=repeated-symbol]{\mega\byte\per\minute}.} Hence, the load $\rho_{obs}^i$ of \gls{ap} $i$ during an observation window $T_{obs}$ is the result of the sum of individual connection contributions:
\begin{equation}
    \rho_{obs}^i = \sum_j \sum_{t \in T_{obs}} \frac{\rho^j}{T^j}\mathbbm{1}_{\{\rho^j_t > 0\}}.
    \label{eq:load}
\end{equation}

\subsection{ML models}

Next, we introduce the state-of-the-art ML models considered for addressing the load prediction problem.

\subsubsection*{ARIMA} ARIMA($p,q,d$) is a statistical model used to predict the evolution of time series through auto-regression ($p$), integration ($q$), and moving average ($d$). In this work, we use ARIMA as a non-ML baseline model. Since ARIMA is a univariate statistical model (it cannot be trained like the rest of ML models and then used to generate predictions on new data), we generate an instance of ARIMA for each AP under evaluation and apply it directly to the evaluation data. For every set of predictions, we fit an ARIMA model with data from the previous 24 hours and apply it using the parameters $p$, $q$, and $d$ obtained through a grid search. This approach is costly and, thus unsuitable to be applied in real-time.

\subsubsection*{LSTM}
A type of \gls{rnn} designed to address long-range dependencies in sequential data thanks to the usage of LSTM cells, which allow capturing temporal dependencies across time series data. In this work, the time series of network data are passed through a few LSTM layers before a predictive output is generated by a feed-forward fully-connected layer. 

\subsubsection*{CNN}
A class of deep NN that automatically learns hierarchical patterns (e.g., edges and shapes from an image) and features from the data by utilizing convolutional layers and other data transformation techniques like down-sampling. Through our solution, the time series vectors of network information features are processed as if they were images by several 2D convolutional layers.

\section{Evaluation}
\label{section:evaluation}

In this section, we evaluate different aspects of the models presented in Section~\ref{section:methodology} when applied to the load prediction problem in enterprise Wi-Fi. For the evaluation of the models, we follow a hold-out validation approach with a sliding window, being the first \qty{80}{\percent} of the measurements of a specific \gls{ap} used as training and validation data and the remaining \qty{20}{\percent} as test data.

\subsection{Data collection requirements}

\begin{figure}[t!]
    \centering    
    \includegraphics[width=\columnwidth]{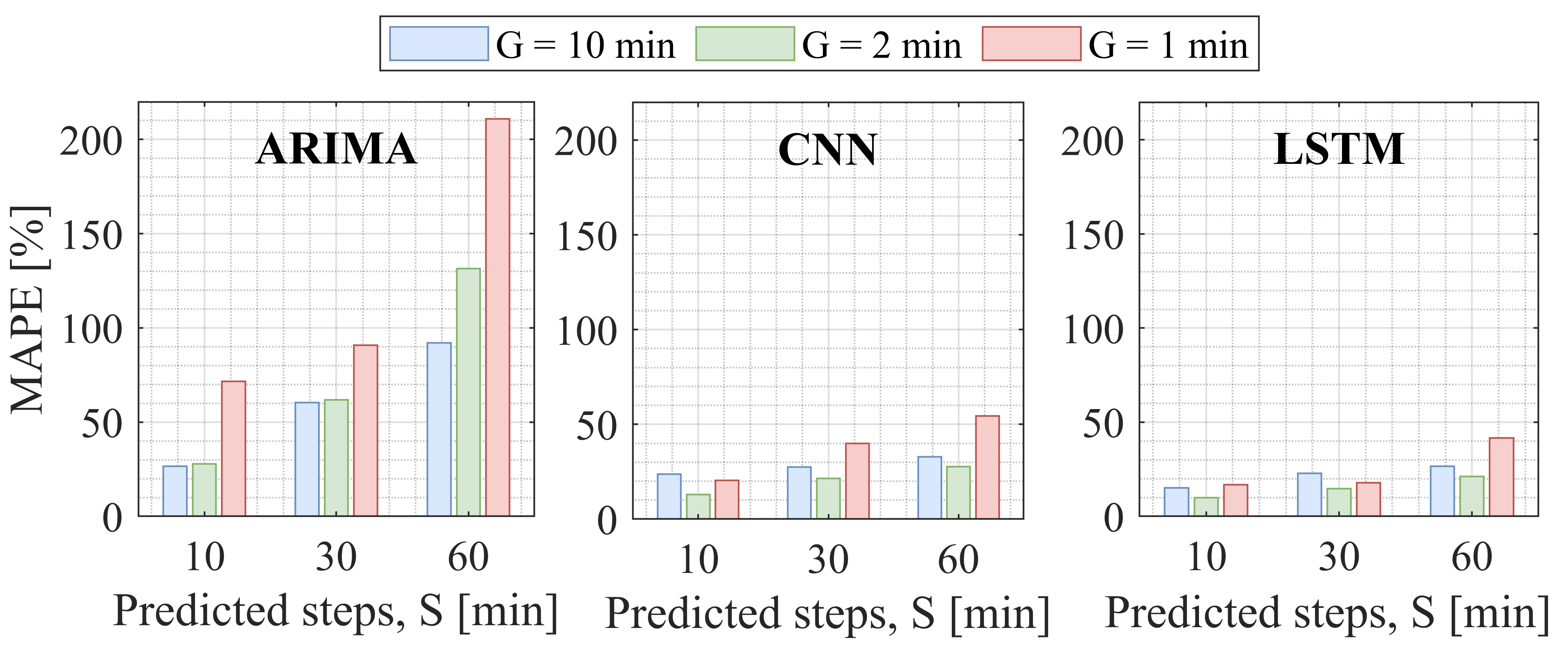}
    \caption{Impact of the granularity of the measurements on the ML models' performance. Predictions for the next $S\in \{\qty{10}{\minute}, \qty{30}{\minute}, \qty{60}{\minute}\}$ are provided for the granularity values $G\in\{\qty{10}{\minute}, \qty{2}{\minute}, \qty{1}{\minute}\}$.}
    \label{fig:data_collection_needs}
\end{figure}

The first critical aspect that comes into play is related to the pace at which data is collected and ingested by ML models. In this regard, the design and size of ML models are strongly influenced by the length of the observation window and the granularity of the measurements.\footnote{For instance, an observation window of \qty{60}{\minute} when using a granularity of \qty{2}{\minute} is equivalent to the acquisition of $\frac{\qty{60}{\minute} }{\qty{2}{\minute} } = 30$ measurements.} The election of the granularity is limited by several factors (e.g., data collection procedure in the network, available memory), and at the same time has a high impact on the accuracy of the models. To showcase the impact of using different granularity values, Fig.~\ref{fig:data_collection_needs} shows the \gls{mape} achieved at $S\in \{\qty{10}{\minute}, \qty{30}{\minute}, \qty{60}{\minute}\}$ steps by each model when using $G\in\{\qty{10}{\minute}, \qty{2}{\minute}, \qty{1}{\minute}\}$ granularity values. To carry out this experiment, we randomly selected a set of \glspl{ap} $\mathcal{K}$ (being $|\mathcal{K}|=16$) and evaluated the solution on the validation data from the same set.

As shown in Fig.~\ref{fig:data_collection_needs}, the granularity of the measurements has a significant impact on the models' performance and poses a trade-off between accuracy and data collection. In particular, \gls{arima} with $G = \qty{10}{\minute}$ leads to prediction errors around 25\% when $S = \qty{10}{\minute}$, increasing to 92\% for farther away predictions ($S = \qty{60}{\minute}$). For \gls{lstm} and \gls{cnn}, in contrast, the prediction error with $G = \qty{10}{\minute}$ remains stable, and mostly below 30\%, at the different values of $S$. 

When it comes to higher granularity of data acquisition (i.e., smaller values of $G$), we find the following. First, $G = \qty{2}{\minute}$ offers the best performance in most of the cases, and this is closely related to the nature of the dataset in use~\cite{chen2021flag} since a significant portion of the connections are below \qty{5}{\minute}. In contrast, $G = \qty{1}{\minute}$ may lead to higher error than for $G = \qty{10}{\minute}$. While higher granularity values allow capturing fast-varying processes, an excessively high granularity may lead to overfitting due to the noisy signals at small scales. For that reason, it is very important to find a good compromise between capturing fine-grained events and keeping the complexity low, and this is something properly captured by $G = \qty{2}{\minute}$ for the particular dataset in use. Based on these observations, we fix the granularity of data acquisition to $\qty{2}{\minute}$ for the remainder of the paper.

\subsection{Training needs and generalization capabilities}

Other important considerations when deploying ML models are related to their training needs and generalization capabilities, i.e., how good are models with unseen data? To assess both properties, we evaluate the different models into different sets of \glspl{ap} with sizes $K_\text{test} \in \{4, 16, 32\}$ and we differentiate between on-premises (\textit{on-prem}) and off-premises (\textit{off-prem}) training approaches. On-prem uses the training data from the same set of test \glspl{ap} ($\mathcal{K}_\text{train}^\text{on-prem} = \mathcal{K}_\text{test}$), which is useful to generate specialized models with potentially superior performance, but it requires significant time for collecting data. As for off-prem training, models are trained using data from a different set of \glspl{ap} ($\mathcal{K}_\text{train}^\text{off-prem} \notin \mathcal{K}_\text{test}$, where $|\mathcal{K}_\text{train}^\text{off-prem}| = 64$ \glspl{ap}), thus becoming less specialized but allowing for immediate deployment of ready-to-use models. Recall that \gls{arima} is directly applied to the set of evaluated \glspl{ap} $\mathcal{K}_\text{test}$ by fitting an \gls{arima} model to each \gls{ap}'s data.

\begin{figure}[t!]
    \centering    \includegraphics[width=0.9\columnwidth]{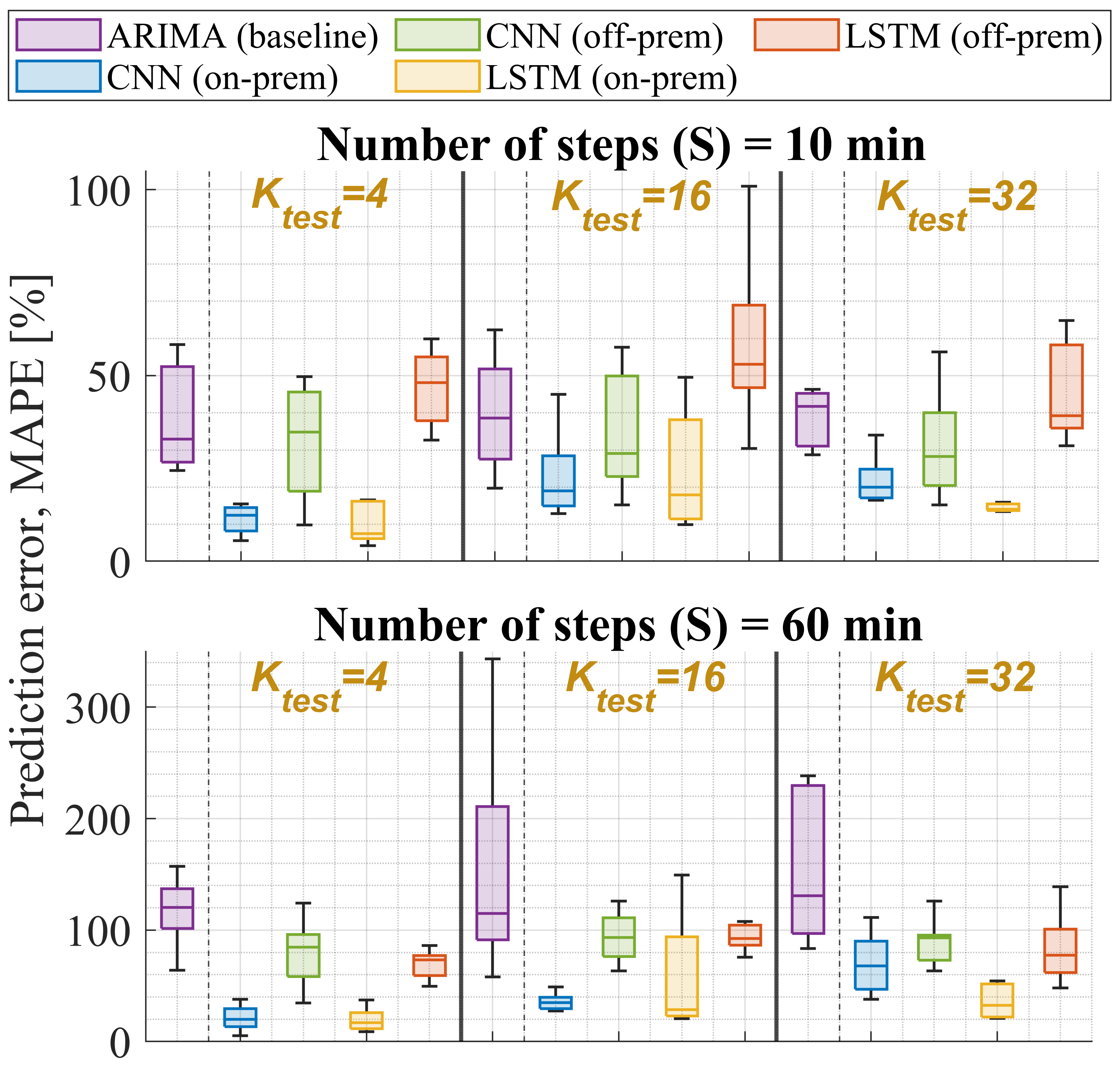}
    \caption{\gls{mape} achieved by the models on test sets corresponding to $K_\text{test} = \{4, 16, 32\}$ \glspl{ap}, for $S\in\{\qty{10}{\minute}, \qty{60}{\minute}\}$.} %
    \label{fig:training_needs}
\end{figure}

Fig.~\ref{fig:training_needs} shows the \gls{mape} achieved through $N=10$ experiments and for $G = \qty{2}{\minute}$. By comparing the performance of the on-prem and off-prem versions of CNN and LSTM, we observe a clear advantage when training models on-prem (at least \qty{10}{\percent} less median error compared to off-prem). On-prem training leads to specialized models with a deeper knowledge of the network in which they are deployed. This effect is intensified in small deployments ($K_\text{test} = 4$), where data heterogeneity is lower, thus allowing for better specialization of the model. As a downside, on-prem training requires doing measurements on-site (e.g., during weeks), which is costly and not always feasible. 

In this respect, we observe that the adoption of off-prem trained models is feasible for short-term predictions ($S = \qty{10}{\minute}$), as this leads to errors around \qty{20}{\percent} and below \qty{50}{\percent} in the first (the bottom line of the box) and third (the top line of the box) quartiles. When targeting further away predictions ($S = \qty{60}{\minute}$), off-prem models lead to very high \gls{mape} values, sometimes above \qty{100}{\percent}. While on one side off-prem models would speed up the deployment of ML solutions at new sites, on the other side considering a short re-training on-premises may be required to decrease their prediction errors.

\begin{figure}[t!]
    \centering    \includegraphics[width=\columnwidth]{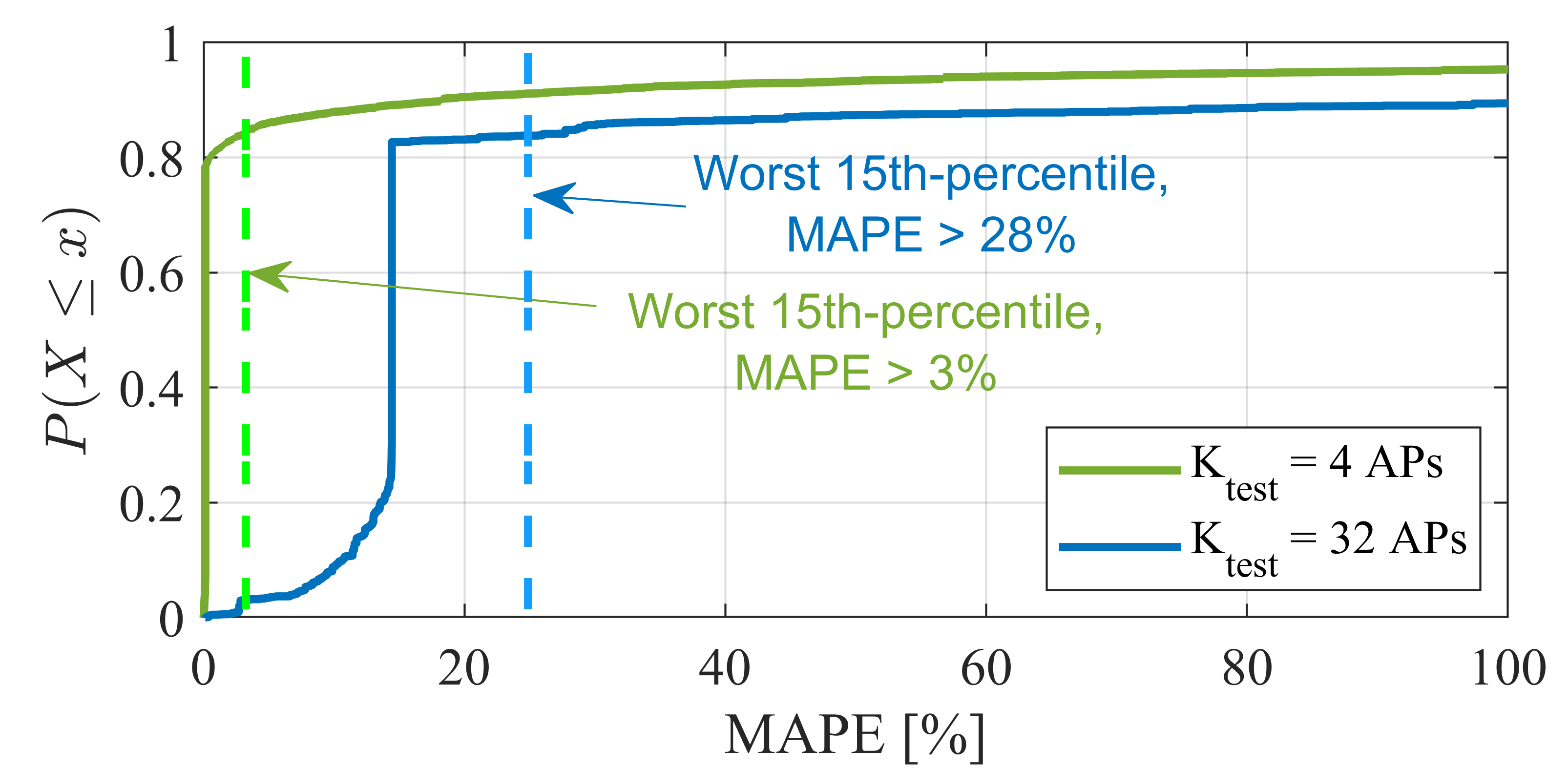}
    \caption{\gls{cdf} of the \gls{mape} obtained by the off-prem \gls{lstm} models when evaluated on $K_\text{test} = \{4, 32\}$ \glspl{ap}, for $S=\qty{10}{\minute}$.}
    \label{fig:cdf_error}
\end{figure}

Given the potential of off-prem models, Fig.~\ref{fig:cdf_error} provides further insights on their performance by showing the \gls{cdf} of the \gls{mape} achieved \gls{lstm} at $K_\text{test} = \{4, 32\}$ \glspl{ap}. As shown, despite leading to a relatively high average error, the model returns very low prediction error occurrences most of the time. Specifically, \qty{85}{\percent} of the predictions present a MAPE under \qty{3}{\percent} and \qty{28}{\percent} for $K_\text{test} = 4$ and $K_\text{test} = 32$, respectively. However, the remaining part of the \gls{cdf} indicates higher experienced errors, mostly triggered by the presence of very high and unexpected peaks of load in the dataset. Although the application of appropriate post-processing techniques would have been beneficial to mitigate the effect produced by these outliers and thus dramatically reduced the MAPE, we decided to not apply any of them to better focus on the capability and performance of the pure ML models.

\subsection{Deployment feasibility and energy consumption}

To assess the feasibility and cost of adopting the proposed ML-based solutions in real deployments, we focus on the energy consumption, the training and inference times, and the memory footprint of the models (shown in Table~\ref{tab:cost_models}). We consider that model (re)trainig is done every two weeks from the data of $K_\text{train}=4$ \glspl{ap}, while model inference is performed every time a new measurement arrives (i.e., every \qty{2}{\minute}) and for each of the evaluated \glspl{ap} ($K_\text{eval}=4$ ). This analysis is conducted in two different computation environments: 1) top-range ML specialized equipment using an A5000 24 GB NVIDIA GPU, and 2) consumer-level equipment using an AMD Ryzen Threadripper 3970X Processor of $\times$32 cores.

As shown, \gls{cnn} and \gls{lstm} lead to similar values, requiring training times of a few seconds and inference times in the order of milliseconds. Moreover, when it comes to memory footprint, the current implementation of the models is not demanding. However, further improvements both in the models (with increased complexity) and the input data (with further features) may easily lead to much more stringent memory requirements, thus requiring specialized hardware like GPUs. Finally, the energy consumed by either applying (inference) and training the models has resulted in general very low, with the inference presenting two orders of magnitude higher values with respect to the training ones due to the higher frequency at which the inference is performed. Overall, this confirms how reasonably good predictions can be obtained even with cost-effective models.

\begin{table}[t]
    \centering
    \caption{Training and inference (in parentheses) cost for the different solutions in terms of computation time, memory footprint, and energy consumption.}
    \resizebox{\textwidth}{!}{%
    \begin{tabular}{@{}lcccc@{}}
    \toprule
    \multicolumn{1}{c}{\textbf{Model}} & \textbf{Env.} & \textbf{\begin{tabular}[c]{@{}c@{}}Comp. \\ time {[}s{]}\end{tabular}} & \textbf{\begin{tabular}[c]{@{}c@{}}Memory$^a$ \\ {[}GPU/CPU \%{]}\end{tabular}} & \textbf{\begin{tabular}[c]{@{}c@{}}Energy$^b$ \\ {[}kWh/month{]}\end{tabular}} \\ \midrule
    \multirow{2}{*}{\textbf{\gls{cnn}}} & \textit{\#1} & 5 (0.00013) & 5.47 (5.4) & 0.0099 (0.5) \\
     & \textit{\#2} & 9.0 (0.0014) & 0.4 (0.3) & 0.0066 (0.17) \\
    \midrule
    \multirow{2}{*}{\textbf{\gls{lstm}}} & \textit{\#1} & 4.3 (0.0009) & 4.95 (4.1) & 0.0062 (0.48) \\
     & \textit{\#2} & 6.8 (0.0012) & 0.4 (0.3) & 0.0046 (0.17) \\
    \midrule
    \end{tabular}%
    }
    \footnotesize{$^a$Measured directly on the device, $^b$Measured using Eco2ai~\cite{budennyy2022eco2ai}.}\\
    \label{tab:cost_models}
\end{table}

\subsection{Model Optimization} 

The utilization of ML models in production environments involves repeating processes like model (re)training and deployment. For this reason, ML model optimization is of utmost importance for the sake of minimizing the operative costs of \gls{aiml}-based applications. Among several model optimization techniques, ML model compression (e.g., pruning, quantization) stands as one of the most appealing methods to save computational time, bandwidth, and energy without sacrificing performance excessively. Although not presented graphically, our results confirmed that when applying post-training static quantization~\cite{nahshan2021loss} to the ML models studied in this paper, their size was reduced by \qty{75}{\percent}, thus speeding up the inference time \numrange{1.5}{1.9} times, while preserving the same accuracy levels.

\section{Conclusions}
\label{section:conclusions}

\gls{aiml}-based network optimization is disrupting enterprise and industrial Wi-Fi solutions, thus becoming an appealing differentiator. A prominent use case of \gls{aiml} in communication is traffic prediction, which unlocks advanced network capabilities for self-troubleshooting or proactively saving energy. In this work, we explored the potential of \gls{aiml}-based traffic prediction in enterprise Wi-Fi networks, linking performance with feasibility. We have delved into data collection requirements, model generalization behaviors, and the cost of adopting simple, but effective, CNN and LSTM models. Our results showed that finding a suitable measurement granularity is key to keeping mean prediction errors below 25\%, as it allows for capturing important events in the network. In addition, we have shown the necessity of using on-prem trained models for use cases requiring far-future predictions, whereas off-prem models could be used instead for short-term predictions with the advantage of being ready to use since their initial deployment, possibly further improved by considering successive on-prem re-training. In terms of cost, the real-time usage of all the studied ML models leads to modest energy consumption values when deployed in commonly available hardware equipment. Regardless of the approach, ML model optimization is a must for \gls{aiml} solutions in production environments, as it leads to substantial savings that can contribute to the sustainable use of ML.


\bibliographystyle{ieeetr}
\bibliography{references}

\begin{thebibliography}{10}

\bibitem{galati2023will}
L.~Galati~Giordano, G.~Geraci, M.~Carrascosa, and B.~Bellalta, ``{What will
  Wi-Fi 8 be? A primer on IEEE 802.11 bn ultra high reliability},'' {\em arXiv
  e-prints}, pp.~arXiv--2303, 2023.

\bibitem{khorsandi2022hexa}
B.~M. Khorsandi, M.~Hoffmann, M.~Uusitalo, {\em et~al.}, ``{Hexa-X deliverable
  D1. 3: Targets and requirements for 6G-initial E2E architecture},'' {\em
  Online: http://hexax. eu/deliverables}, 2022.

\bibitem{troia2018deep}
S.~Troia, R.~Alvizu, Y.~Zhou, G.~Maier, and A.~Pattavina, ``{Deep
  learning-based traffic prediction for network optimization},'' in {\em 2018
  20th International Conference on Transparent Optical Networks (ICTON)},
  pp.~1--4, IEEE, 2018.

\bibitem{aiml_tig}
{IEEE 802.11 Working Group (WG)}, ``{11-22/0597r3: 802.11 May 2022 WG
  Motions},'' 2022.

\bibitem{wfa_dataelem}
{Wi-Fi Alliance}, ``{ Wi-Fi CERTIFIED Data Elements™ (Version 2.1)},'' 2022.

\bibitem{tedjopurnomo2020survey}
D.~A. Tedjopurnomo, Z.~Bao, B.~Zheng, F.~M. Choudhury, and A.~K. Qin, ``{A
  survey on modern deep neural network for traffic prediction: Trends, methods
  and challenges},'' {\em IEEE Transactions on Knowledge and Data Engineering},
  vol.~34, no.~4, pp.~1544--1561, 2020.

\bibitem{abbasi2021deep}
M.~Abbasi, A.~Shahraki, and A.~Taherkordi, ``{Deep learning for network traffic
  monitoring and analysis (NTMA): A survey},'' {\em Computer Communications},
  vol.~170, pp.~19--41, 2021.

\bibitem{chen2021flag}
W.~Chen, F.~Lyu, F.~Wu, P.~Yang, and J.~Ren, ``{Flag: Flexible, accurate, and
  long-time user load prediction in large-scale WiFi system using deep RNN},''
  {\em IEEE Internet of Things Journal}, vol.~8, no.~22, pp.~16510--16521,
  2021.

\bibitem{kirchgassner2012introduction}
G.~Kirchg{\"a}ssner, J.~Wolters, and U.~Hassler, {\em Introduction to modern
  time series analysis}.
\newblock Springer Science \& Business Media, 2012.

\bibitem{hernandez2009arima}
C.~A. Hern{\'a}ndez~Suarez, O.~J. Salcedo~Parra, and A.~Escobar~D{\'\i}az,
  ``{An ARIMA model for forecasting Wi-Fi data network traffic values},'' {\em
  Ingenier{\'\i}a e Investigaci{\'o}n}, vol.~29, no.~2, pp.~65--69, 2009.

\bibitem{jin2012characterizing}
Y.~Jin, N.~Duffield, A.~Gerber, P.~Haffner, W.-L. Hsu, G.~Jacobson, S.~Sen,
  S.~Venkataraman, and Z.-L. Zhang, ``Characterizing data usage patterns in a
  large cellular network,'' in {\em Proceedings of the 2012 ACM SIGCOMM
  workshop on Cellular networks: operations, challenges, and future design},
  pp.~7--12, 2012.

\bibitem{wilhelmi2020flexible}
F.~Wilhelmi, S.~Barrachina-Mu{\~n}oz, B.~Bellalta, C.~Cano, A.~Jonsson, and
  V.~Ram, ``{A flexible machine-learning-aware architecture for future
  WLANs},'' {\em IEEE Communications Magazine}, vol.~58, no.~3, pp.~25--31,
  2020.

\bibitem{szott2022wi}
S.~Szott, K.~Kosek-Szott, P.~Gaw{\l}owicz, J.~T. G{\'o}mez, B.~Bellalta,
  A.~Zubow, and F.~Dressler, ``{Wi-Fi meets ML: A survey on improving IEEE
  802.11 performance with machine learning},'' {\em IEEE Communications Surveys
  \& Tutorials}, vol.~24, no.~3, pp.~1843--1893, 2022.

\bibitem{feng2006svm}
H.~Feng, Y.~Shu, S.~Wang, and M.~Ma, ``{SVM-based models for predicting WLAN
  traffic},'' in {\em 2006 IEEE international conference on communications},
  vol.~2, pp.~597--602, IEEE, 2006.

\bibitem{thapaliya2018predicting}
A.~Thapaliya, J.~Schnebly, and S.~Sengupta, ``{Predicting congestion level in
  wireless networks using an integrated approach of supervised and unsupervised
  learning},'' in {\em 2018 9th IEEE Annual Ubiquitous Computing, Electronics
  \& Mobile Communication Conference (UEMCON)}, pp.~977--982, IEEE, 2018.

\bibitem{khan2020real}
M.~A. Khan, R.~Hamila, N.~A. Al-Emadi, S.~Kiranyaz, and M.~Gabbouj,
  ``{Real-time throughput prediction for cognitive Wi-Fi networks},'' {\em
  Journal of Network and Computer Applications}, vol.~150, p.~102499, 2020.

\bibitem{jiang2022cellular}
W.~Jiang, ``Cellular traffic prediction with machine learning: A survey,'' {\em
  Expert Systems with Applications}, vol.~201, p.~117163, 2022.

\bibitem{trinh2018mobile}
H.~D. Trinh, L.~Giupponi, and P.~Dini, ``Mobile traffic prediction from raw
  data using lstm networks,'' in {\em 2018 IEEE 29th annual international
  symposium on personal, indoor and mobile radio communications (PIMRC)},
  pp.~1827--1832, IEEE, 2018.

\bibitem{albawi2017understanding}
S.~Albawi, T.~A. Mohammed, and S.~Al-Zawi, ``Understanding of a convolutional
  neural network,'' in {\em 2017 international conference on engineering and
  technology (ICET)}, pp.~1--6, Ieee, 2017.

\bibitem{zhang2020citywide}
D.~Zhang, L.~Liu, C.~Xie, B.~Yang, and Q.~Liu, ``Citywide cellular traffic
  prediction based on a hybrid spatiotemporal network,'' {\em Algorithms},
  vol.~13, no.~1, p.~20, 2020.

\bibitem{feng2018deeptp}
J.~Feng, X.~Chen, R.~Gao, M.~Zeng, and Y.~Li, ``{Deeptp: An end-to-end neural
  network for mobile cellular traffic prediction},'' {\em IEEE Network},
  vol.~32, no.~6, pp.~108--115, 2018.

\bibitem{perifanis2023towards}
V.~Perifanis, N.~Pavlidis, S.~F. Yilmaz, F.~Wilhelmi, E.~Guerra, M.~Miozzo,
  P.~S. Efraimidis, P.~Dini, and R.-A. Koutsiamanis, ``{Towards Energy-Aware
  Federated Traffic Prediction for Cellular Networks},'' in {\em 1st
  International Symposium on Federated Learning Technologies and Applications
  (FLTA)}, 2023.

\bibitem{budennyy2022eco2ai}
S.~A. Budennyy, V.~D. Lazarev, N.~N. Zakharenko, A.~N. Korovin, O.~Plosskaya,
  D.~V. Dimitrov, V.~Akhripkin, I.~Pavlov, I.~V. Oseledets, I.~S. Barsola, {\em
  et~al.}, ``{Eco2ai: carbon emissions tracking of machine learning models as
  the first step towards sustainable AI},'' in {\em Doklady Mathematics},
  vol.~106, pp.~S118--S128, Springer, 2022.

\bibitem{nahshan2021loss}
Y.~Nahshan, B.~Chmiel, C.~Baskin, E.~Zheltonozhskii, R.~Banner, A.~M.
  Bronstein, and A.~Mendelson, ``Loss aware post-training quantization,'' {\em
  Machine Learning}, vol.~110, no.~11-12, pp.~3245--3262, 2021.

\end{thebibliography}

\end{document}